\documentstyle[aps,prl,multicol,epsf]{revtex}
\begin{document}
\title{\bf Avalanches and Waves in the Abelian Sandpile Model}

\author{Maya Paczuski$^1$ and Stefan Boettcher$^{2}$}
\address{$^1$Department of Physics, University of Houston, Houston TX
77204-5506\\$^2$Center for Nonlinear Studies,
MS-B258, Los Alamos National Laboratory, Los Alamos, NM 87545}
\date{\today}

\maketitle 

\begin{abstract}

We numerically study avalanches in the two dimensional Abelian
sandpile model in terms of a sequence of waves of toppling events.
Priezzhev et al [PRL {\bf 76}, 2093 (1996)] have recently proposed
exact results for the critical exponents in this model based on the
existence of a proposed scaling relation for the difference in sizes
of subsequent waves, $\Delta s =s_{k}- s_{k+1}$, where the size of the
previous wave $s_{k}$ was considered to be almost always an upper
bound for the size of the next wave $s_{k+1}$.  Here we show that the
significant contribution to $\Delta s$ comes from waves that violate
the bound; the average $\langle\Delta s(s_{k})\rangle$ is actually
negative and diverges with the system size, contradicting the proposed
solution.

\end{abstract}


\begin{multicols}{2}

The  sandpile  was the first model
 introduced by Bak, Tang, and Wiesenfeld to demonstrate the
 principle of self-organized criticality  \cite{BTW}.
 Self-organized criticality describes a general
 property of slowly-driven dissipative systems with many degrees of
 freedom to evolve toward a stationary state where activity takes
 place intermittently in terms of bursts spanning all scales up to the
 system size.  The sandpile
  model has subsequently received a great deal of attention
 due in part to its potential for having a theoretical solution.  Dhar
 showed that certain aspects of its behavior could be calculated
 exactly based on the Abelian symmetry of topplings.
  For instance, rigorous results have been obtained for the total number 
  of allowed configurations on the attractor of recurrent states \cite{Dhar1}, 
  for some height-height correlation functions \cite{M+D,Ivash}, and later for 
  the distribution of sizes of the last wave in an
 avalanche \cite{D+M}, among other quantities.  Nevertheless, a formal solution
 for the all important critical exponents describing the distribution
 of avalanche sizes and durations in the Abelian sandpile model
 (ASM) has remained an elusive goal.

In a recent Letter \cite{PKI} exact results were proposed for the
distribution of avalanche sizes in the ASM, based on a decomposition of
 an avalanche into a sequence of ``waves'' of topplings.
Specifically,  the prediction was made that
 the  asymptotic distribution of the number
of topplings in an avalanche is $P(S) \sim S^{-\tau}$ with $\tau=6/5$,
and the distribution of the number of sites covered by an avalanche is
$P(a) \sim a^{-\tau_{a}}$ with $\tau_{a}=5/4$. The results of
numerical simulations do not convincingly support these claims,
although measuring the actual avalanche distribution exponents in this model is
notoriously difficult.

 Here we scrutinize the main assumption in the argument leading to the
 predicted exact results in Ref.~\cite{PKI} for the
 ASM.  Based on careful numerical simulations we show that the
 fundamental assumption that the next wave usually is contained by 
the previous wave fails drastically.  In
 particular the exponent $\alpha$ as defined in \cite{PKI} does not
 exist, and the difference in sizes of subsequent waves $\Delta s$ is more
often negative than positive.  In fact, the negative contribution
 has a sufficiently fat tail that the average difference $\langle
 \Delta s(s_{k})\rangle$ is negative and diverges with system size
for all $s_{k}< s_{co}$.
  The quantity $s_{co}$ is the cutoff in waves sizes
 due to the finite size system, $s_{co}\sim L^{2}$, where $L$ is the
 linear extent of the system.
 In short, the physics
 is dominated by waves that are not contained within their
 predecessors.  Thus the argument leading to the claimed exact results
 is incorrect.

The ASM consists of a square lattice of size $L$ with a discrete
number $z_i$ of sand grains occupying each site. Initially, the
lattice may be empty, and sand is dropped grain by grain at random
sites. After each drop, all sites which exceed a critical threshold
for stability, $z_i>z_{\rm c}=3$, are ``toppled'' by
distributing a single grain of sand to each of their four nearest neighbors or,
for boundary sites, over the edge of the lattice. Toppling proceeds
for a number of time steps until all sites are stable again, and a
new grain is dropped at a random site. Dropping sand represents an
external driving force on the system whose impact is dissipated in
intermittent sequences of toppling events which are called
avalanches. The number of toppling events following the addition of a
single grain is the size $S$ of an avalanche.
Starting from an empty lattice, avalanches are initially
rare and only of short duration. But the system fills up with
sand to the point that many sites are close to threshold. Then, the
system  reaches a stationary state in which for any one
grain dropped, on average, one grain must leave the system through the open
boundaries. The grains are transported by avalanches which are now
broadly distributed in both duration and extent over many orders of magnitude,
only limited by the system's size. Thus, the system has
self-organized into a critical (SOC) state with a highly correlated
response to the external driving.

This model has a few other remarkable properties, most notable the
fact that the order of toppling events during an avalanche is
interchangeable (``Abelian'') without changing the final state of the
system, which for instance allows an exact enumeration of the critical
state \cite{Dhar1}.  Also, it was found that the domain spanned by a
single avalanche is always compact, though with a fractal boundary
\cite{C+O}.  Manna \cite{manna} introduced a different sandpile model, 
without Abelian symmetry, where  the toppling grains
are stochastically distributed to nearest neighbors so that the
distribution is symmetric only on average. For some time it was believed,
 based on real space renormalization group arguments \cite{pietronero}
 and Manna's numerical simulation results, that the ASM was in the same
universality class as the Manna model. But for both models the values of 
the distribution exponents, obtained by various extrapolations of the 
results from extensive numerical simulations, have barely converged to 
within 10\% after many years of study.  Recently,  Ben-Hur and Biham 
\cite{biham} computed the geometrical scaling properties of avalanches 
instead of individual avalanche distribution exponents. Their results 
indicate that the ASM and the
Manna model may belong to different universality classes. A survey of two-time
autocorrelation functions in various SOC models \cite{SocAg1} also reveals
differences between both models: the ASM exhibits ``aging'' \cite{aging}
while the Manna model does not \cite{SocAg1,ASMage}.

Effort has focused recently on understanding   avalanche
dynamics in the ASM by decomposing the avalanche into a sequence of
more elementary events.  Dhar and Manna \cite{D+M} introduced the notion of
inverse avalanches which were shown to be equivalent to a direct
representation of avalanches in terms of waves of toppling events
\cite{IKP}.  Ivashkevich, Ktitarev, and Priezzhev defined waves as
follows: if the site (i) to which a grain was added becomes unstable,
topple it once and then topple all other sites that become unstable,
keeping the initial site (i) from toppling a second time.  The set of
sites that toppled thus far are called ``the first wave of
topplings'' since every site can only topple once.  After the first wave is
 completed the site (i) is
allowed to topple the second time, not permitting it to topple again
until the ``second wave of topplings'' is finished.  The process
continues until the site (i) becomes stable and the avalanche stops.

This elegant decomposition of avalanches into waves reveals many
interesting features.  First of all, the waves are individually
compact, and each
site that topples in a wave topples exactly once in that wave.  As a
result, the
state of the system after a wave is exactly the same as the state
before the wave except at sites on the single closed boundary of the
wave which separates sites that toppled in that wave from sites that did not.
Just inside  the wave boundary a trough relative to the previous heights
appears, whereas just outside the boundary a hill appears where sand
was transport outside of the wave.  Thus the sequence of topplings in
the next wave will follow exactly the sequence of topplings in the
previous wave until the wave first reaches the prior wave's boundary, at which
point the sequence may differ.  Priezzhev et al
\cite{PKI} argued that generally subsequent
waves are spatially contained within the previous waves because of the
trough at the boundary.  Their analysis only considers waves
which are contained within previous waves and they ``neglect the
overlapping of waves and deal only with the decrease of wave size''
\cite{PKI}.  Using spanning tree
arguments together with this assumption, they find that the size
difference between subsequent waves $<\Delta s> = s_{k}- s_{k+1}$ is
positive, finite in the limit of large system size,
and obeys a scaling relation $\Delta s \sim s_{k}^{\alpha}$.  Key to
the argument leading to Eq.~(12) in Ref.~\cite{PKI}
is the length of the boundary $\Gamma$ of the previous wave
$s_{k}$ which is presumed to contain the subsequent wave $s_{k+1}$.
 Here we show that the main contribution to $\Delta s$ comes from waves
which escape the boundary of their preceding wave, and are bounded
only by the system size.  In fact, the average
$\langle \Delta s(s_{k}) \rangle$ is negative and diverges to
infinity as the system size increases.

We have simulated about $10^{7}$ waves in $L^2$-systems up to  $L=1024$.
To ensure the accuracy of our numerical simulations, we have
reproduced a variety of previously obtained exact results for the
distribution of waves. For instance, our data yields the $s^{-11/8}$
power-law that was derived by Dhar and Manna \cite{D+M}
for the distribution of the very last wave in each avalanche. We also
found the $1/s$  behavior for the distribution of all waves \cite{IKP}.

\begin{figure}
\narrowtext
\epsfxsize=2.2truein
\hskip 0.15truein\epsffile{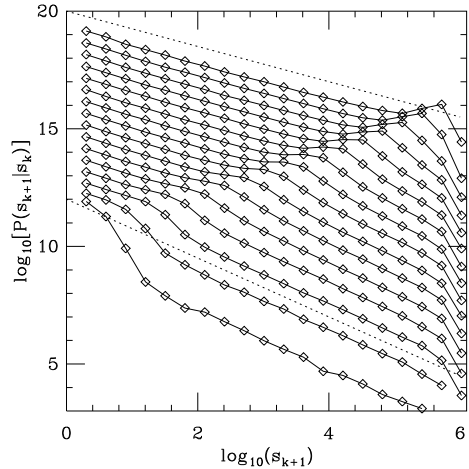}
\caption{
Plot of the distribution $P(s_{k+1}|s_{k})$ for the
next wave to be of size $s_{k+1}$, given that the previous wave was
of size $s_{k}$ in a system of size $L=1024$. 
Each graph contains data for $2^{m}\leq s_{k}<
2^{m+1}$ for $m=4,5,\ldots,19$ from bottom to the top. To avoid
overlaps, the graphs are offset. It appears that for each value of
$s_{k}$,  $P(s_{k+1}|s_{k})$ initially falls like $s_{k+1}^{-3/4}$ (as
the dashed line on top),
but at a scale linear in $s_{k}$ crosses over to a $s_{k+1}^{-5/4}$
fall-off (as the lower dashed line).
}
\end{figure}

To determine $\Delta s$ we recorded the distribution of subsequent waves
 for a given size of the preceding wave $P(s_{k+1}|s_{k})$ as shown in Fig.~1.
 In Fig.~2 we show the data collapse where the horizontal axis is now
$x= s_{k+1}/s_{k}$.  There are two regimes separated by a turning point
near one.  There is clear evidence of a fat power law tail for
$x \gg 1$, which will dominate the average $\Delta s$ for each value of
$s_{k}$.
 The data is sufficiently well represented by a scaling
form
\begin{equation}
P(s_{k+1}|s_{k}) \sim s_{k+1}^{-\beta}F\left({s_{k+1}\over
s_{k}}\right) \quad ,
\label{scal}
\end{equation}
where $F(x\rightarrow 0) \rightarrow 1$ and $F (x \gg 1) \sim
x^{-r}$.  The data collapse indicates that $\beta \simeq 3/4$ and $ r \simeq
1/2$.

\begin{figure}
\narrowtext
\epsfxsize=2.2truein
\hskip 0.15truein\epsffile{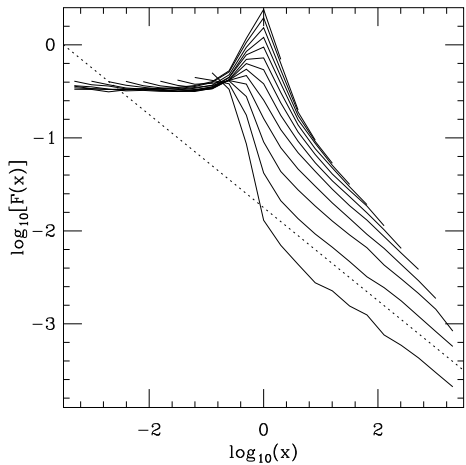}
\caption{
Scaling collapse for the data in Fig.~1 according to
Eq.~(\protect\ref{scal}), giving $F(x)\sim
s_{k+1}^{3\over 4}P(s_{k+1}|s_{k})$ as a function of $x=s_{k+1}/s_{k}$.
The tail for each graph falls approximately like $x^{-{1\over 2}}$ (as 
the dashed line).
}
\end{figure}

{}From Eq.~(\ref{scal}), the computation of
$\langle \Delta s(s_{k}) \rangle = <s_{k}- s_{k+1}>$
leads immediately to
\begin{equation}
\langle \Delta s(s_{k}) \rangle
= s_{k}- s_{k}\int^{s_{co}/s_{k}} dx~x^{1-\beta}F(x)\quad ,
\label{dsint}
\end{equation}
where $s_{co}$ is the cutoff in wave sizes from the finite
system size; $s_{co} \sim L^{2}$.
 For $s_{k}\ll s_{co}$ this gives $<\Delta s(s_{k})>=s_{k} - C
 s_{co}^{2-\beta -r}s_{k}^{\beta +r -1}$ or, using our values for
$\beta$ and $r$,
\begin{eqnarray}
<\Delta s(s_{k})>= s_{k} - C s_{co}^{3/4}s_{k}^{1/4}\quad,
\label{ds}
\end{eqnarray}
where $C$ is a positive number that  depends on the details of the function
$F$.  {\it It is important to note here that $\langle \Delta s \rangle$
is negative for all $s_{k}$ up to a scale $L^{2}$, and that it diverges
 with the system size.} Our numerical
measurement for $\langle \Delta s \rangle$
shown in Fig.~3 confirms the above analysis.

If we (erroneously) exclude from the data set all waves that are larger than
 their predecessor, i. e. eliminating those waves that contribute to a
negative $\Delta s$, then we reproduce (in Fig.~4) the plot given in  Fig.~1
of Ref.~\cite{PKI}.

\begin{figure}
\narrowtext
\epsfxsize=2.2truein
\hskip 0.15truein\epsffile{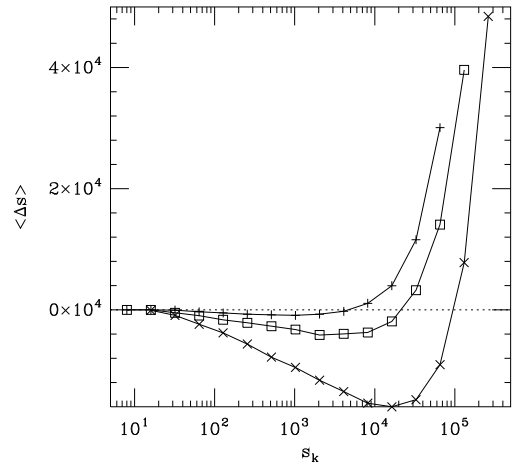}
\caption{
Plot of $<\Delta s>=s_k-s_{k+1}$ as a function of the previous wave $s_k$,
obtained from $P(s_{k+1}|s_k)$ (see Fig.~1) via Eq.~(\protect\ref{dsint})
for system size $L=2^8,~2^9$ and $2^{10}$ from top to bottom.
Initially, $<\Delta s>$ is negative and falling for increasing $s$.
Closer to the cut-off $s_{co}$ the linearly rising term in
Eq.~(\protect\ref{ds}) dominates. In each case, the graph turns
positive at $s_k\approx L^2/10$.
}
\end{figure}
\begin{figure}
\narrowtext
\epsfxsize=2.2truein
\hskip 0.15truein\epsffile{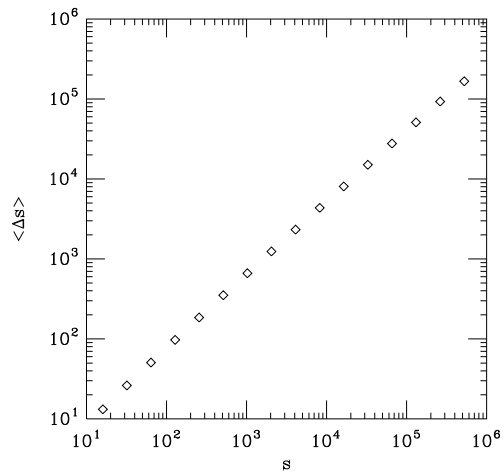}
\caption{
Plot of $<\Delta s>$ as a function of the previous wave $s$,
but leaving out all data where $\Delta s$ is negative. We chose $L=512$ 
to compare with Fig.~1 in Ref.~\protect\cite{PKI}. 
}
\end{figure}

{}Furthermore, a more detailed analysis of subsequent waves that {\it
are} smaller, in terms of number of topplings $s$,
 than their predecessor reveals that in most cases
the following (smaller) wave still, more often than not,
exceeds the confines of the
(larger) previous wave at some points on the boundary.  The smaller
waves actually do escape the boundary of their predecessor.
 As the system size grows, the
the fraction of consecutive waves which violate the assumption
of Ref.~\cite{PKI} also grows.

In summary, we have shown that the analysis in Ref. \cite{PKI} is
fundamentally flawed because it deals only with waves of decreasing
size, whereas the dominant contribution to $\Delta s$ comes from the
``fat tails'' in the distribution $P(s_{k+1}|s_{k})$ where waves escape
the boundary of their predecessor explicitly. Yet, our numerical data
seems to indicate that a certain degree of regularity in the distribution of
consecutive waves exists, leading to what appears to be exact values for
$\beta=3/4$ and $r=1/2$.  It may be possible that the spanning tree
arguments used in Ref. \cite{PKI} can be generalized to include the
dominant contribution from overlapping waves.  The appearance of the
apparently simple exponents $\beta$ and $r$ gives us some hope that an
exact solution using the elegant decomposition
of avalanches into waves can be discovered.

Both of us gratefully acknowledge the hospitality of the Santa Fe
Institute where part of this work was completed, and STB thanks the
Physics Department at the University of Houston for its hospitality.

\end{multicols}
\end{document}